# Dose-response curve of EBT, EBT2, and EBT3 radiochromic films to synchrotron-produced monochromatic x-ray beams.

Thomas A. D. Brown, Kenneth R. Hogstrom

*Mary Bird Perkins Cancer Center, 4950 Essen Lane, Baton Rouge, LA 70809 and Department of Physics and Astronomy, Louisiana State University and A & M College, 202 Nicholson Hall, Baton Rouge, LA 70803*

Diane Alvarez, Kenneth L. Matthews II

*Department of Physics and Astronomy, Louisiana State University and A & M College, 202 Nicholson Hall, Baton Rouge, LA 70803*

Kyungmin Ham

*Center for Advanced Microstructures and Devices, Louisiana State University and A & M College, 6980 Jefferson Highway, Baton Rouge, LA 70806*

Joseph P. Dugas

*Mary Bird Perkins Cancer Center, 4950 Essen Lane, Baton Rouge, LA 70809 and Department of Physics and Astronomy, Louisiana State University and A & M College, 202 Nicholson Hall, Baton Rouge, LA 70803*

**ABSTRACT**

**Purpose:** This work investigates the dose-response curves of GAFCHROMIC® EBT, EBT2, and EBT3 radiochromic films using synchrotron-produced monochromatic x-ray beams. EBT2 film is being utilized for dose verification in photoactivated Auger electron therapy at the Louisiana State University Center for Advanced Microstructures and Devices (CAMD) synchrotron facility.

**Methods:** Monochromatic beams of 25, 30, and 35 keV were generated on the tomography beamline at CAMD. Ion chamber depth-dose measurements were used to determine the dose delivered to films irradiated at depths from 0.7 to 8.5 cm in a 10×10×10-cm$^3$ polymethylmethacrylate phantom. AAPM TG-61 protocol was applied to convert measured ionization into dose. Films were digitized using an Epson 1680 Professional flatbed scanner and analyzed using the net optical density (NOD) derived from the red channel. A dose-response curve was obtained at 35 keV for EBT film, and at 25, 30, and 35 keV for EBT2 and EBT3 films. Calibrations of films for 4 MV x-rays were obtained for comparison using a radiotherapy accelerator at Mary Bird Perkins Cancer Center.

**Results:** The sensitivity (NOD per unit dose) of EBT film at 35 keV relative to that for 4-MV x-rays was 0.73 and 0.76 for doses 50 and 100 cGy, respectively. The sensitivity of EBT2 film at 25, 30, and 35 keV relative to that for 4-MV x-rays varied from 1.09 – 1.07, 1.23 – 1.17, and 1.27 – 1.19 for doses 50 – 200 cGy, respectively. For EBT3 film the relative sensitivity was within 3% of unity for all three monochromatic x-ray beams.

**Conclusions:** EBT and EBT2 film sensitivity showed strong energy dependence over an energy range of 25 keV – 4 MV, although this dependence becomes weaker for larger doses. EBT3 film shows weak energy dependence, indicating that it would be a better dosimeter for kV x-ray beams where beam hardening effects can result in large changes in the effective energy.

## I. INTRODUCTION

GAFCHROMIC® EBT2 radiochromic film (International Specialty Products, Wayne, NJ) serves as a secondary dosimeter in clinical radiation therapy. The properties and sensitivity of EBT2 film have been well documented for a wide range of x-ray beam energies.[1-6] Although the dose response of EBT2 film has been shown to have minimal energy dependence in the megavoltage energy range, there are data to indicate variation with energy for kilovoltage x-rays.[1-5] Data provided by the manufacturer indicates that the film sensitivity may be as much as 10% higher for kV x-rays compared to MV x-rays.[1] Butson et al. observed a variation in sensitivity of up to 7% for energies 50 – 250 kVp with a maximum sensitivity 5% greater than MV energies at 100 kVp.[2] A broad variation in sensitivity (± 20 %) with chemical composition was seen by Lindsay et al. at 105 kVp, and they reported energy independence from 105 kVp – 6 MV for the film composition adopted since May 2009.[3] More recently, Arjomandy et al. reported dose-response curves for kV and MV x-rays, $^{60}$Co gamma, $^{137}$Cs gamma, electron, and proton beams used in radiotherapy.[4,5] The film sensitivity to 75 kVp x-rays was found to be approximately 9% higher compared to photons in the MV energy range. They also showed a change in dose response versus depth in water at 75 kVp due to beam hardening, another indication of the variation of sensitivity with energy.

All of these results for photons were reported for polychromatic x-ray beams or gamma rays. EBT2 film is now being used for dosimetry measurements of low-energy monochromatic x-ray beams (25 – 35 keV) produced at the Louisiana State University Center for Advanced Microstructures and Devices (CAMD) synchrotron facility. These measurements are being used to verify the dose delivered to cells undergoing photoactivated Auger electron therapy.[7] It is therefore important to understand the response of this film to low-energy monochromatic x-rays. The successor to EBT2 film, EBT3, was made commercially available in September 2011 and may be a more suitable dosimeter for low-energy x-rays. The manufacturer reports that the composition of the active layer remains unchanged[8], and a recent study of the film indicates that it has similar dosimetric properties to EBT2 film.[9] However, the new film differs from EBT2 in that it is symmetric (the outer polyester layers are of equal thickness) so that the film may be scanned on either side, and there are microscopic silica particles embedded into the polyester substrate to prevent the formation of Newton's rings in images obtained using a flatbed scanner.[8]

In earlier work at CAMD, EBT film was used to verify dose for cell irradiations performed at 35 keV.[7] Changes in the chemical composition between lots of EBT film manufactured at different times has led to substantial differences in the energy response.[3] The energy dependence of the film used for this earlier work was investigated by measuring dose-response curves for x-ray beams at 6 MV, 125, and 75 kVp.[10] The film sensitivity at 200 cGy, measured relative to 6 MV, was found to be 0.76 at 75 kVp and 0.81 at 125 kVp. The calibration obtained using the 125-kVp beam was used to describe the dose response at 35 keV since they had the most similar Al half-value layer (HVL) values (3.0mm vs. 3.3mm, respectively). There were no measurements made to determine a dose-response curve for EBT film directly from the monochromatic beam; however, the dose derived using the 125-kVp calibration curve was found to be only 3% lower than that determined from ion chamber measurements at 35 keV.[11]

This paper describes the calibration and energy dependence of EBT2 and EBT3 films using monochromatic beams at 25, 30, and 35 keV. The results are described relative to the response from a 4-MV calibration obtained using a clinical radiotherapy accelerator. For the purposes of comparison to previous results, new calibrations of EBT film at 35 keV and 4 MV are also presented.

## II. METHODS AND MATERIALS

**II.A Monochromatic x-ray source**

Monochromatic x-ray beams of 25, 30, and 35 keV were generated on the tomography beamline at the CAMD synchrotron facility. A 1.3-GeV electron beam ($I_{max}$ = 220 mA) was transported through a three-pole superconducting wiggler magnet ($B_{max}$=7T), creating a polychromatic x-ray beam. Monochromatic x-rays ($\Delta E/E \approx 2$ %) were selected by transporting the beam through a calibrated W-$B_4C$ double-multilayer monochromator (Oxford Danfysik, UK). Due to physical restrictions imposed by the monochromator and the beamline slits, the resulting monochromatic beam was approximately 3.0-cm wide × 0.2-cm high. The narrow beam was filtered using 640 µm Al since low-energy x-ray contamination can be significant. The energy of the beam was verified using the Debye-Scherrer cones produced from Si640c powder diffraction.[11] A flat-panel XRD 0820 CN3 detector (PerkinElmer, Waltham, MA) was used to measure the resulting diffraction rings, allowing for energy precision to be within 0.1 keV. An effective broad beam approximately 3.0-cm wide × 2.5-cm high was created by vertically oscillating the irradiation target through the path of the narrow beam at 0.125 cms$^{-1}$. Target oscillation was achieved using a screw-drive motion stage (Velmex, Inc., Bloomfield, NY) controlled by a user-programmed LabVIEW (National Instruments Corporation, Austin, TX) interface. Previous measurements have shown that the effective broad beam can be considered parallel.[11] Conventional, 4-MV x-rays were produced using a Clinac 21EX radiotherapy accelerator (Varian Medical Systems, Inc., Palo Alto, CA) at Mary Bird Perkins Cancer Center.

**II.B Dose calibration**

Ionization chamber depth-dose measurements of the monochromatic beams were used to calibrate the films. The dose delivered by the beam in a 10×10×10-cm$^3$, 1.18 gcm$^{-3}$, polymethylmethacrylate (PMMA) phantom was measured using a calibrated 0.23-cm$^3$ Scanditronix Wellhofer model FC23-C cylindrical, air-equivalent ion chamber (Scanditronix Wellhofer GmbH, Schwarzenbruck, Germany) with a Modified Keithley 614 Electrometer (CNMC Company, Best Medical, Nashville, TN). The ion chamber was used to measure the ionization created by the effective broad beam along its central axis for phantom depths from 0.6 to 10.0 cm. The time to irradiate with a broad beam was specified in terms of the number of complete stage oscillations, and this ensured that the dose delivery was uniform in the vertical direction. Each irradiation measurement was conducted for 320 s, corresponding to eight complete stage oscillations. The x-ray dose output was proportional to the ring current which slowly decayed between electron injections into the synchrotron storage ring (over ~

7 hours). Using the average ring current for each irradiation the measured ionization was normalized to a ring current of 100 mA.

The AAPM TG-61 protocol[12] for determining dose to water ($D_w$) for medium-energy x-rays (100 kV – 300 kV) at 2-cm depth, was applied to convert the ring-current normalized ionization ($M_{norm}$) at all depths into dose:

$$D_w = M_{norm} P_{elec} P_{TP} P_{ion} P_{pol} P_{Q,cham} N_k \left(\frac{\mu_{en}}{\rho}\right)_{AIR}^{WATER}, \tag{1}$$

where $P_{elec}$ is the electrometer accuracy correction factor, $P_{TP}$ is the ambient temperature and pressure correction factor, $P_{ion}$ is the ion recombination correction factor, $P_{pol}$ is the polarity effect correction factor, $P_{Q,cham}$ is the overall chamber correction factor, $N_k$ is the air-kerma calibration factor of the ion chamber, and $\left(\frac{\mu_{en}}{\rho}\right)_{AIR}^{WATER}$ is the ratio of the water-to-air mass-energy absorption coefficients. The "in-air" dose calibration method used for low- and medium-energy x-rays (40 kV – 300 kV) was equally applicable to this work. However, the "in-phantom" method was chosen because of the lack of low-energy monochromatic data available for backscatter factors required for the "in-air" method.

The ion chamber correction and calibration factors were obtained in the same way as described by Oves et al.[10] and are shown for each energy in Table 1. The ion chamber measurements used to calculate $P_{ion}$ and $P_{pol}$ were conducted at a PMMA depth of 0.6 cm using the same broad beam geometry as the depth-dose measurements. Irradiations were typically performed for 160 s (four stage oscillations), and the measured ionization was normalized to a ring current of 100 mA. $P_{ion}$ was determined for the case of a continuous beam using high and low electrometer bias voltages of -300 and -150 V, respectively. Values for $P_{Q,cham}$ were difficult to determine since the energies and field size used for these measurements lay outside the range of data available for this correction factor in TG-61.

Estimates of $P_{Q,cham}$ were obtained by using $P_{Q,cham} = 0.995$ for the similar NE2611/NE2561 chambers and for a 0.1 mm Cu HVL beam in TG-61 Table VIII, and then applying a field size correction factor of 1.005 by extrapolating data in TG-61 Figure 4 for the broad beam size (7.5 cm$^2$) used in this work. $N_k$ was determined using a linear fit to ADCL calibrated values measured for a 120 kVp beam (HVL=6.96 mm Al) and an 80 kVp beam (HVL = 2.96 mm Al), which were $1.215 \times 10^8$ Gy C$^{-1}$ and $1.219 \times 10^8$ Gy C$^{-1}$, respectively. The HVL values were used to interpolate/extrapolate $N_k$ values at 35 keV (HVL = 3.33 mm Al), 30 keV (HVL = 2.28 mm Al), and 25 keV (HVL = 1.12 mm Al). Mass-energy absorption coefficients were interpolated for each energy using NIST tables[13] and used to calculate values for $\left(\frac{\mu_{en}}{\rho}\right)_{AIR}^{WATER}$.

| TG-61 factor | 25 keV | 30 keV | 35 keV |
|---|---|---|---|
| $P_{elec}$ | 0.987 | 0.987 | 0.987 |
| $P_{TP}$ | 1.000 – 1.021 | 1.001 – 1.002 | 1.003 – 1.011 |
| $P_{ion}$ | 1.000 – 1.002 | 1.000 | 1.000 – 1.008 |
| $P_{pol}$ | 0.994 – 0.999 | 0.995 | 0.990 – 1.003 |
| $P_{Q,cham}$ | 1.000 | 1.000 | 1.000 |
| $N_k$ | $1.221 \times 10^8$ Gy C$^{-1}$ | $1.220 \times 10^8$ Gy C$^{-1}$ | $1.219 \times 10^8$ Gy C$^{-1}$ |
| $\left(\frac{\mu_{en}}{\rho}\right)_{AIR}^{WATER}$ | 1.019 | 1.013 | 1.015 |

TABLE 1: TG-61 ion chamber calibration and correction factors used for dose calculations at 25, 30, and 35 keV. Measurements of $P_{TP}$, $P_{ion}$ and $P_{pol}$ were repeated for each set of depth-dose measurements and the range of values obtained are shown here (only one set of $P_{ion}$ and $P_{pol}$ measurements were made at 30 keV).

The principle source of uncertainty in the normalized ionization values arose from small variations in the beam output that were independent of the ring current. These variations can arise as a result of changes in the phase space of the ring electrons, beamline vacuum fluctuations, or beam heating of the monochromator. The standard deviation of multiple normalized ionization values measured at a single PMMA depth was used to estimate the size of this uncertainty. This uncertainty was propagated to determine the uncertainty in $P_{ion}$ and $P_{pol}$. The total uncertainty in the corrected, normalized, ionization value ($M_{norm}P_{elec}P_{TP}P_{ion}P_{pol}P_{Q,cham}$) used to determine the TG-61 dose was found by propagating the uncertainty in $P_{ion}$, $P_{pol}$, and $M_{norm}$, and was determined to be ± 3%. The accuracy of the ion-chamber measured dose has been assessed by converting measured incident fluence into dose using MCNP5 calculations of dose per fluence.[11,14] In the most recent work[14], the fluence-based dose overestimated the ion-chamber measured dose at 25 keV by an average of 7.2 ± 3.0% to 2.1 ± 3.0% for PMMA depths from 0.6 to 7.7 cm, respectively. At 35 keV, the fluence-based dose underestimated the ion chamber measurements by an average of 1.0 ± 3.4% to 2.5 ± 3.4%, respectively. Based on these results, an uncertainty of ±5% (1σ) was adopted for the total uncertainty in the TG-61 measured dose at energies 25 – 35 keV. This value includes the uncertainty associated with the beam output variations and the systematic error associated with the TG-61 calibration factors derived for this work.

**II.C Film irradiations**

The films studied in this work were obtained from lots # 48022-05 (EBT), A02181103 (EBT2), and A09231103 (EBT3). For each calibration, up to thirteen $5 \times 5$ cm$^2$ pieces of film were cut from a single sheet with a small line drawn on each piece to indicate the orientation of the original. Two pieces of film were used to provide a background measurement of the optical density for each sheet of film. All of the EBT3 film used for this work was obtained from a single $35.6 \times 43.2$ cm$^2$ sheet.

The ion chamber measurements were used to determine the dose delivered by the monochromatic beams at different depths in the PMMA phantom. Dose-response measurements were performed by irradiating eight $5 \times 5$ cm$^2$ pieces of film within the phantom at depths from 0.7 to 8.5 cm. The PMMA phantom was composed of 0.7 and 1.2-cm thick $10 \times 10$ cm$^2$ plates, and the film pieces were sandwiched between these plates for irradiation. Each film piece was centered laterally and taped to an adjacent plate. The plates were then aligned and taped together to minimize the effect of air gaps in the phantom. EBT film pieces were irradiated at each depth individually while each set of EBT2 and EBT3 films were irradiated at all depths simultaneously. The front surface of each piece of film was used as the effective point of measurement (film thickness <0.3 mm), and the depth values were determined accordingly. Small corrections (<0.2 cm) to the EBT2 and EBT3 film depths were made to account for the presence of film pieces at shallower depths in the phantom by calculating the increased beam attenuation. These depth corrections are included in the range of depths given above. The length of each irradiation was chosen so that a dose of ~ 200 cGy was delivered at a depth of 0.7 cm. The time and average ring current for each film irradiation were used to renormalize the measured ion chamber dose output. A 4$^{th}$-order polynomial fit to the ion chamber depth-dose measurements was used to interpolate the dose for each piece of film at 35 keV, and 5$^{th}$-order fits were used at 25 and 30 keV. The reproducibility of this calibration method was tested by repeating the calibrations of EBT and EBT2 films. The dose-response curve was measured three times at 35 keV for EBT film and twice at 25 and 35 keV for EBT2 film. A new set of ion-chamber measurements were made for each calibration. The difference in film response for the repeated calibrations was consistent with the uncertainty associated with the variation in beam output discussed in Section II B. These results indicated that we were able to achieve a high level of consistency for each calibration setup.

For the 4-MV irradiations, between eight and eleven $5 \times 5$ cm$^2$ pieces of film were irradiated individually at 90-cm SSD with a $30 \times 30$-cm$^2$ field, defined at isocenter, at a 10-cm depth in a Solid Water® (GAMMEX rmi, Middleton, WI, USA) phantom. Doses to the film were determined using standard monitor unit calculations based on dose output calibrated at 100-cm SAD following TG-51 protocol.[15] X-ray dose output constancy from the radiotherapy accelerator is checked daily at Mary Bird Perkins Cancer Center in accordance with TG-40 protocol.[16] The output tolerance for daily checks is prescribed as 3%; therefore, an uncertainty of ±2% was adopted for the delivered film doses.

**II.D Film digitization and analysis**

All irradiated films were digitized using an Epson 1680 Professional flatbed scanner (Seiko Epson Corporation, Nagano, Japan) at least 24 hours after irradiation. Andres *et al.* reports that the post-exposure optical density of EBT and EBT2 film stabilizes in as little as 6 and 2 hours, respectively.[6] To avoid systematic errors arising from film handling, each piece of film was cleaned with a 70% ethanol solution prior to scanning to remove any finger prints and any remaining felt pen marks (other than the orientation line) used to divide up the original sheet of film. The film, centered on the scanner bed using a cardboard template to ensure film placement reproducibility[17], was aligned so that the long axis of the scanner was parallel to the long axis of the film[6]. Due to the asymmetric structure of EBT2 film[1], care was taken to ensure that the film was always scanned with the 50 μm polyester layer facing the glass window on the bed of the scanner. To avoid any large change in light intensity as the scanner lamp warmed up, five scans were initially performed in the absence of film to ensure a stable light output.[17] The film was scanned in transmission mode using the software Image Acquisition (International Specialty Products, Wayne, NJ) and was stored as a 300 dpi, 16-bit, TIFF image.

The TIFF images were analyzed using the software ImageJ v1.42q (National Institutes of Health, Bethesda, MD). For the monochromatic irradiations, the exposed region of each piece of film consisted of a 3.0 cm × 2.5 cm area centered on the film. The mean pixel value for the red channel was obtained for a region of interest (ROI) measuring 0.5 cm × 0.5 cm centered on the film image. The center of the film corresponded to the effective point of measurement for the ion chamber. For the 4-MV irradiations, the entire area of the film was uniformly irradiated and therefore a larger region of interest (~ 1.5 cm × 1.5 cm) was selected. The unexposed pieces of film were analyzed in the same way. The red channel pixel values were converted into optical density using a $5^{th}$-order polynomial calibration curve obtained from a NIST calibrated TIFFEN Transmission Photographic Step Tablet #2 (The Tiffen Company, Rochester, NY). For each calibration, the net optical density was calculated by taking the difference between the optical density of the exposed film and the average optical density determined from the two unexposed pieces of film.

The uncertainty in the net optical density was derived from three sources of error associated with the mean pixel value for each ROI: (1) changes in the scanner lamp output after warm-up (2) variation in pixel value across the ROI due to statistics and beam non-uniformities and (3) differences in the background optical density for each $5 \times 5$ cm$^2$ piece of film. Changes in the scanner lamp output were investigated by performing consecutive scans of a single piece of film. The mean pixel value for both the small and larger ROIs discussed above was found to vary by less than ±0.06% (1σ) after completion of the lamp warm-up procedure. The standard error in the mean was used to estimate the uncertainty associated with the variation in pixel value across the ROI. This error was very small (less than ±0.01%) due to the large number of pixels associated with each ROI. Scans of unexposed $5 \times 5$ cm$^2$ pieces of film, obtained from the same sheet, were used to determine the effect of variations in the background optical density. After subtracting the effect of lamp output variations, the mean pixel value was found to vary by less than ±0.2%, corresponding to a variation of less than ±0.8% in the background optical density. Error propagation of these uncertainties gave rise to a net optical density uncertainty of less than ±1%.

Net optical density was plotted versus dose for each calibration, and a function of the form recommended by the manufacturer[1] was fitted to the data using a non-linear least squares algorithm in Gnuplot v4.4 (www.gnuplot.info):

$$NOD = -\ln\left(\frac{a+b.D}{a+D}\right). \qquad (2)$$

**III. RESULTS AND DISCUSSION**

Figs. 1 – 3 show plots of the net optical density versus dose for EBT, EBT2, and EBT3 film. The calibration function given in Eq. (2) has been fitted to all the available data for each beam energy. The relative sensitivity of each type of film from 50 – 200 cGy, determined from these fitted curves, is quantified in Table 2.

| Beam energy (keV) | Dose (cGy) | EBT | EBT2 | EBT3 |
|---|---|---|---|---|
| | | **Relative sensitivity:** $\frac{Net\ optical\ density(E)}{Net\ optical\ density(4\ MV)}$ | | |
| 25 | 50 | - | 1.09 | 0.97 |
| | 100 | - | 1.08 | 0.97 |
| | 200 | - | 1.07 | 0.97 |
| 30 | 50 | - | 1.23 | 0.98 |
| | 100 | - | 1.21 | 0.99 |
| | 200 | - | 1.17 | 0.99 |
| 35 | 50 | 0.73 | 1.27 | 0.99 |
| | 100 | 0.76 | 1.24 | 0.98 |
| | 200 | - | 1.19 | 0.97 |

TABLE 2: Comparison of film sensitivity with beam energy for doses 50, 100, and 200 cGy. The relative sensitivity was calculated using net optical density values determined from the fitted calibration function given in Eq. (2).

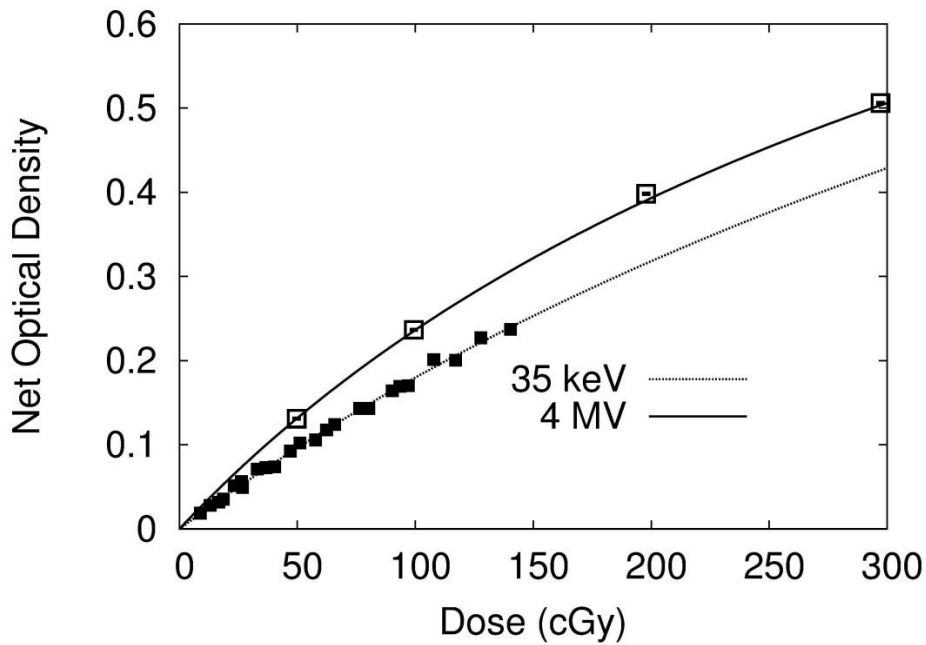

FIG. 1: Net optical density versus dose for EBT film. The film was calibrated at 35 keV (filled squares) and 4 MV (hollow squares). Both sets of data were fitted with the function shown in Eq. (2).

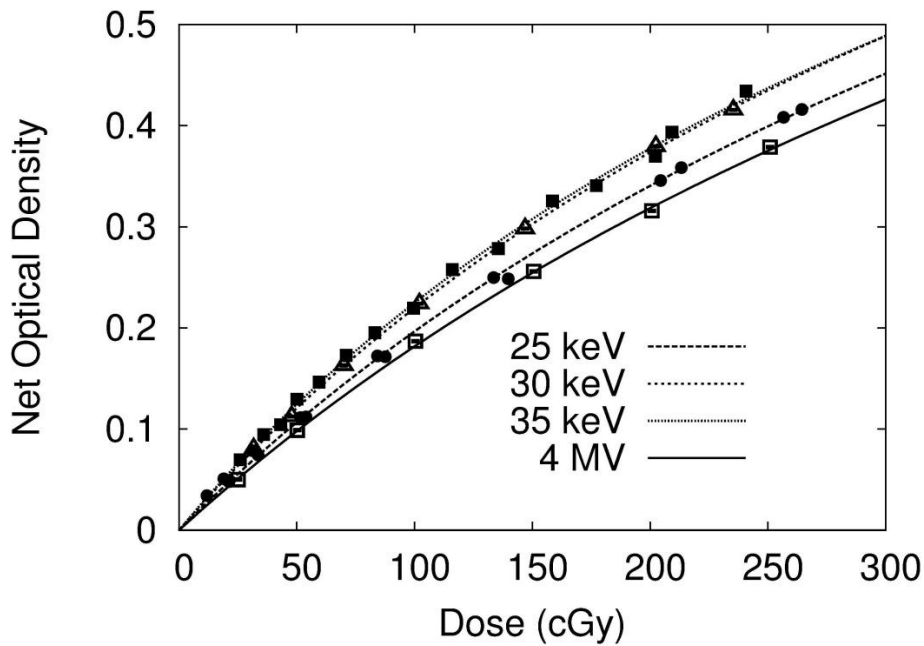

FIG. 2: Net optical density versus dose for EBT2 film. The film was calibrated at 25 keV (circles), 30 keV (triangles), 35 keV (filled squares) and 4 MV (hollow squares). Each set of data were fitted with the function shown in Eq. (2).

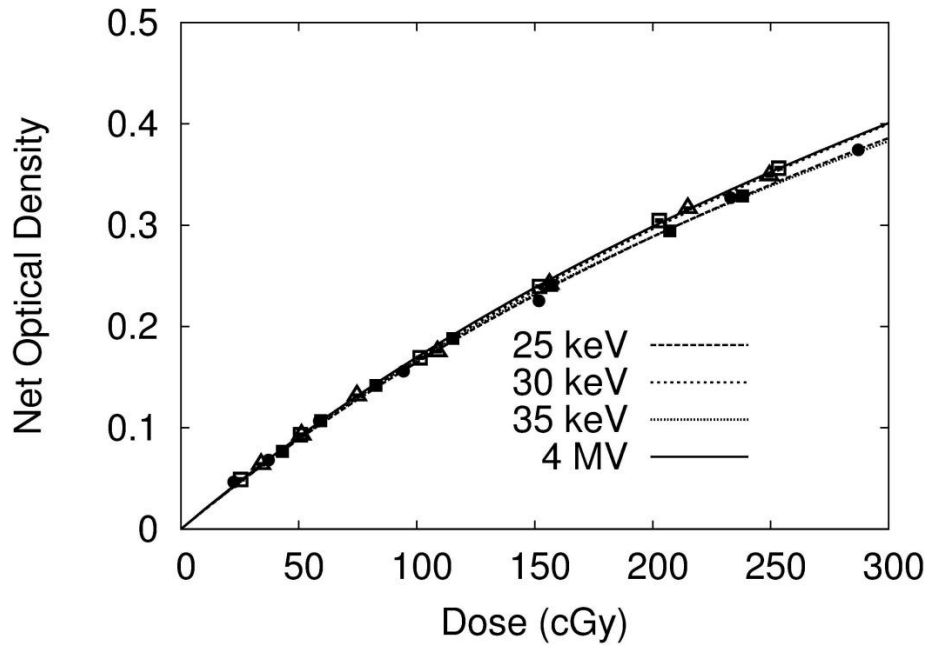

FIG. 3: Net optical density versus dose for EBT3 film. The film was calibrated at 25 keV (circles), 30 keV (triangles), 35 keV (filled squares) and 4 MV (hollow squares). Each set of data were fitted with the function shown in Eq. (2).

The uncertainties associated with the NOD and the dose were not used to fit the data. All of the data was fitted using a non-weighted least squares method. Notwithstanding this simplistic approach, this analysis is robust enough to provide a reliable comparison of the film sensitivity from 25 − 35 keV. The uncertainty in the dose consists of two components: (1) beam output variations independent of the storage ring current, and (2) the systematic error associated with the TG-61 calibration factors. The output variations can give rise to scatter in the film response for a given dose, although this effect is partly mitigated for the EBT film and the EBT2 film at 25 and 35 keV, where the function has been fitted to data consisting of two or more separate calibrations. The systematic error is the dominant source of uncertainty in the dose and is common to all the measurements at 25 − 35 keV. Thus, while the absolute values for the relative sensitivities given in Table 2 may be slightly offset from their true values, their magnitude relative to one another is sufficiently insensitive to the uncertainty in the dose to provide a reliable intercomparison. Weighted fits, using the uncertainties in the NOD values, were performed and found to produce no significant difference in the relative sensitivity results.

Arjomandy *et al.* conducted depth-dose measurements with EBT2 film using 75 kVp x-rays and reported a small variation in film sensitivity with depth that was attributed to beam hardening.[5] An over-response of 0.9% relative to ion chamber measurements was observed after an effective beam energy change of 6.3 keV. Although the films for this work were calibrated using monochromatic beams, there was high-energy contamination that arose as a result of the transmission of photons through the monochromator satisfying the Bragg condition ($n\lambda=2d\sin\theta$) for n=2. X-ray scatter measurements of the beam using a NaI detector indicated that <4% of the photons satisfy this condition.[14] Monte Carlo simulations of beam transport through the PMMA phantom[11,14] indicated that the mean photon energy increases by <1 keV at 35 keV and <3 keV at 25 keV as the beam penetrates from 0.7 to 8.5 cm. This data indicated that beam hardening should have had no significant effect on our results.

The sensitivity (NOD per unit dose) of EBT film from 50 – 150 cGy is significantly lower at 35 keV compared with 4 MV. The relative sensitivity increases by 4% over this dose range possibly indicating weaker energy dependence at higher doses. The change in sensitivity with energy is consistent with results reported around 100 kVp for the more recent lots of EBT film, which showed relative sensitivities increasing from 0.68 – 0.81 for 50 – 200 cGy.[3,10,18] The decrease in energy dependence with dose has been explained by Lindsay *et al.*[3] (for EBT2 as well as EBT) in terms of a simple physical model.

EBT2 film shows a higher sensitivity from 50 – 200 cGy for all three monochromatic beams compared with 4 MV. In each case the relative sensitivity decreases with dose, particularly at 30 and 35 keV, indicating weaker energy dependence at higher doses. The energy dependence varies significantly between 25 and 35 keV. The sensitivity peaks at 30 – 35 keV and is lower at 25 keV. The higher relative sensitivity of EBT2 film at 25 – 35 keV is broadly consistent with data reported in the literature,[1,2,4,5] although the values reported in this study are generally higher. It's likely that this difference is due to a combination of the uncertainty in the dose and the use of monochromatic x-rays in this work. It's also possible that small differences in the high-Z composition of film from different lot numbers could account for sensitivity differences seen among researchers.[3] The energy dependence of the EBT2 film from 25 – 35 keV is qualitatively similar to the results reported by Butson *et al.*[2] Their study reported average relative sensitivities of 1.05, 1.04, and 0.985 for doses 50 – 200 cGy at 100 kVp ($E_{eq}$ = 36 keV), 75 kVp ($E_{eq}$ = 30 keV), and 50 kVp ($E_{eq}$ = 25.2 keV), respectively. The results from this study show average relative sensitivities of 1.23, 1.20, and 1.08 at 35, 30, and 25 keV, respectively.

In contrast with EBT and EBT2 film, the sensitivity of EBT3 film shows weak energy dependence between 25 keV and 4 MV. The relative sensitivity from 50 – 200 cGy agrees to within 3% of unity for all three beam energies. This result was unexpected given that the manufacturer has reported no difference between the active layers of EBT2 and EBT3 film.[1]

## IV. CONCLUSIONS

EBT and EBT2 film sensitivity show strong energy dependence over an energy range of 25 keV – 4 MV, and EBT2 even shows significant dependence from 25 – 35 keV. The energy dependence of both films becomes weaker for higher doses. EBT3 sensitivity shows a weak energy dependence over an energy range of 25 keV – 4 MV. This work indicates that EBT3 film would be a better dosimeter for kV x-ray beams where beam hardening effects can result in large changes in the effective energy, although researchers should always verify the energy-response characteristics of their batch of film.

ACKNOWLEDGEMENTS

The authors would like to thank David Perrin and Connel Chu (Mary Bird Perkins Cancer Center) for their assistance with the 4-MV irradiations.

This research was supported by contract W81XWH-10-1-0005 awarded by The U.S Army Research Acquisition Activity, 820 Chandler Street, Fort Detrick, MD 21702-5014. This paper does not necessarily reflect the position or policy of the Government, and no official endorsement should be inferred.